\documentstyle[12pt]{article}

\def\a{\alpha}
\def\b{\beta}

\def\d{\delta}
\def\f{\phi}                    
\def\g{\gamma}

\def\j{\psi}
\def\l{\lambda}

\def\o{\omega}
\def\p{\pi}                     
\def\q{\theta}                  
\def\s{\sigma}                  
\def\t{\tau}

\def\F{\Phi}

\def\cd{{\cal D}}

\def\cl{{\cal L}}

\def\co{{\cal O}}

\newfam\BMath
\font\BMathL=cmbx10 at 12pt
\font\BMathl=cmbx10 at 8.5pt
\font\BMathm=cmbx8  at 6pt
\textfont\BMath=\BMathL \scriptfont\BMath=\BMathl
\scriptscriptfont\BMath=\BMathm

\def\intks{\int \frac{\mbox{\rm d}^3 \mbox{\bf k}}{(2\pi)^3}}
\def\intps{\int \frac{\mbox{\rm d}^3 \mbox{\bf p}}{(2\pi)^3}}
\def\intqs{\int \frac{\mbox{\rm d}^3 \mbox{\bf q}}{(2\pi)^3}}
\def\intxs{\int \mbox{\rm d}^3 \mbox{\bf x}}

\def\rd{\mbox{\rm d}}

\def\exp{\mbox{\rm exp}}
\def\tr{\mbox{\rm tr \ }}
\def\Del{\mbox{\boldmath $\partial$}}
\def\del{\mbox{$\partial$}}
\def\ncdot{\!\cdot\!}

\newcommand{\ncom}{\newcommand}
\ncom{\beq}{\begin{equation}}
\ncom{\eeq}{\end{equation}}
\ncom{\beqa}{\begin{eqnarray}}
\ncom{\eeqa}{\end{eqnarray}}
\ncom{\beqano}{\begin{eqnarray*}}
\ncom{\eeqano}{\end{eqnarray*}}
\ncom{\bep}{\begin{picture}}
\ncom{\eep}{\end{picture}}
\ncom{\eref}[1]{Eq. (\ref{#1})}
\ncom{\vo}[1]{{\fam\BMath #1}}
\ncom{\vt}[2]{({\fam\BMath #1}+{\fam\BMath #2})}
\ncom{\vmo}[1]{\vert{\fam\BMath #1}\vert}
\ncom{\vmt}[2]{\vert\mbox{\bf #1}+\mbox{\bf #2}\vert}
\ncom{\hq}{\hat Q}
\ncom{\vhq}{\hat \vo Q}

\ncom{\lan}{\langle}
\ncom{\ran}{\rangle}
\ncom{\fx}{\!\!\!\!}
\ncom{\nonum}{\nonumber \\}
\ncom{\dbd}[1]{\mbox{$\mbox{\rm d} \over {\mbox{\rm d} #1}$}}

\ncom{\klo}{k_{10}}
\ncom{\kto}{k_{20}}
\ncom{\dad}{\Del \!\cdot\! \dot {\vo A}}
\ncom{\da}{\Del \!\cdot\! {\vo A}}
\ncom{\lsta}{\lan T \f|}
\ncom{\rsta}{|T' \f' \ran}
\ncom{\at}{{\vo A}_T}
\ncom{\al}{{\vo A}_L}

\ncom{\half}{{1\over 2}}
\ncom{\third}{{1\over 3}}
\ncom{\fourth}{{1\over 4}}
\ncom{\fifth}{{1\over 5}}
\ncom{\sixth}{{1\over 6}}

\begin{document}

\begin{titlepage}

\begin{centering}
\null
\vspace{1cm}
{\bf DIFFICULTIES IN FORMULATING AN IMAGINARY-TIME}

\vspace{0.3cm}
{\bf FORMALISM OF TEMPORAL AXIAL GAUGE AT FINITE T}

\vspace{2cm}
{\bf S.M.H. Wong}

\vspace{1cm}
\em \footnote{Laboratoire associ\'e au Centre National de la
Recherche Scientifique}LPTHE, Universit\'e Paris-Sud, B\^atiment 211,
F-91405 Orsay, France

\end{centering}

\vspace{3cm}

\begin{center}
{\bf Abstract}
\end{center}

\vspace{0.5cm}

Following the successful construction of the finite T real-time formalism
in temporal axial gauge, we attempt to further study the equivalent
new imaginary-time formalism of James and Landshoff based on the same
Hamiltonian approach in the hope that it will provide the answer to
Debye screening in QCD. It turns out that, unlike in the real-time case,
energy conservation does not hold because of the unusual
representation of the longitudinal field forced upon by the
Hamiltonian formulation.

\vspace{2.0cm}
\noindent LPTHE-Orsay 94/114

\noindent November 1994 (revised)

\addtocounter{page}{1}
\thispagestyle{empty}
\pagebreak
\end{titlepage}

\section{Introduction}

In our previous attempt \cite{pei&wong} to study the screening behaviour of
the singlet quark-antiquark potential in a quark-gluon plasma within
the framework of Braaten and Pisarski resummed perturbation
theory \cite{bra&pis} in temporal axial gauge (TAG),
we encountered some difficulties in connection with
the longitudinal propagator. They arise due to the singular
$1/p_0^2$ factor at $p_0=0$ of the Matsubara frequency sum. These
difficulties were overcome in earlier works \cite{kaj&kap,hei&kaj&toi}
by using a pragmatic ansatz which amounted to assuming the
standard formula \cite{kap} for performing the frequency sum
\beq T\sum_n f(p_0)={1\over {2i\pi}} \Big \{
     \int^{i\infty}_{-i\infty} \rd z f(z) +
     \int^{i\infty+\epsilon}_{-i\infty+\epsilon} \rd z
     \Big (f(z)+f(-z) \Big ) {1 \over {e^{z/T}-1}} \Big \}
\label{ansatz1}
\eeq
holds even in the case that $f(z)$ has poles on the imaginary axis.
So we have the following vanishing sums
\beq T \sum 1/p_0 = 0 \mbox{\rm \hskip1cm , \hskip 1cm}
     T \sum 1/p^2_0 =0 \; .
\label{ansatz2}
\eeq
As noted in these papers, there exists no proof that this is
correct to any order although at leading order this is found to be
correct.

Indeed by using this ansatz, we found, at next to leading order in
TAG, a power screening behaviour which is in contradiction with that obtained
in covariant gauge and Coulomb gauge \cite{reb1,reb2,reb3}. Moreover, it
is also in disagreement with the result obtained from the correlator of two
Polyakov loops \cite{reb4,bra&nie}. However, with the use of this ansatz,
similar result was also found in TAG by some other unique
construction \cite{baier&kal}.

Instead of using such an ansatz, one can use a general axial gauge
$n \ncdot A =0$ with a time-like 4-vector $n^\mu = (1,\vo n)$ and
obtain TAG in the limit $\vo n \rightarrow 0$. However, the
usual simplification of the vanishing space-time, time-time components
of the propagator in TAG is gone \cite{nach}.

It has been pointed out by James and Landshoff \cite{james&land} that
the imaginary-time longitudinal propagator used in the previous works
cannot be correct because including only physical states in the thermal
average breaks the periodicity of the longitudinal propagator. They
have initiated a new formalism in which the longitudinal propagator
comes out automatically free of the $p_0=0$ singularity. They have
shown that one can obtain the same answer of the two-loop pressure
in this formalism as in covariant gauge and Feynman gauge
\cite{james&land,james3}.

In the hope that this formalism will provide the answer to
the above contradiction and difficulties and to see whether the power
screening is a physical result or a gauge artifact, we study this new
formalism. One striking feature of this formalism is that the simple
basic energy conservation mechanism in the usual theory is absent so
one may wonder whether energy conservation is still there. We will
explain this mechanism in the usual theory and why it is absent
in the new formalism in the next section. In section \ref{searchO},
we find the physical states for performing the thermal average and
we will show explicitly the absence of energy conservation
in some one-loop propagators in section \ref{secprop}.
In section \ref{puzzle}, we will give an argument for energy
conservation and in section \ref{explain}, we will point out the
reason of energy non-conservation.

\section{James and Landshoff's TAG Formalism}

\label{jameslandform}

In this section, we describe briefly the formalism of James and
Landshoff \cite{james&land}.

At zero temperature, TAG is known to require a prescription to deal
with the $1/p_0^2$ factor in the longitudinal propagator. It has been
the case that Wilson loop calculation becomes the almost standard test
for any potentially correct prescription. The Vienna prescription
\cite{gai&kreu} which is a generalization of the Mandelstam-Leibbrandt
prescription \cite{lei} in light-cone gauge was found to be better
than the others such as principal value prescription. It is therefore
natural to try to extend this to finite temperature.
This has been done by James in real-time \cite{james1,james2,james3}
and James and Landshoff in imaginary-time \cite{james&land} based
on the Hamiltonian approach to the Vienna prescription derived in
\cite{land}. Here we concern ourselves only with the imaginary-time
formalism.

With the absence of ghost in TAG, the thermal average
has to be taken only over physical states. One immediate
consequence is that the longitudinal propagator is not periodic
in time despite the fact that gluons are bosons (compared with
the usual formalism). This has to do with the fact that whether
the propagator of a given particle is periodic depends
not only on its species but also on whether it is heated in the heat
bath. This peculiar situation happens only in the case of gauge
particles due to the Hilbert space occupied by them is not everywhere
physical. It is easy to see this in a simplified situation of a
free theory. Recall that in quantum mechanics, the amplitude
for evolving from a state $Q$ at $T$ to a state $Q'$ at $T'$
is given by
\beq \lan Q' T' | Q T \ran = \int [dQ] \; \exp \{ -i \int^{T'}_T dt L \} \; ,
\eeq
so one can convert the partition function for, say scalar theory, into
path integral form, i.e.
\beqa Z_\f &=& \tr \lan \f | \; \exp (-\b H_\f) \; | \f \ran  \nonum
           &=& \int_{\rm periodic} [d\f] \;
           \exp \{ -i \int^{-i\b}_0 dt L_\f \}    \; .
\eeqa
The interpretation is that one starts from a state $\f$ at $t=0$
and evolves back through a time $-i\b$ to $\f$ so that $\f$ has to
be periodic in $-i\b$. The equivalent for gauge fields in the free
field case is
\beqa Z_{A} &=& \tr \lan T| \; \exp (-\b H_A) |T \ran  \nonum
            &=& \int_{\rm periodic} [dA_T] \;
            \exp \{ -i \int^{-i\b}_0 dt L_{A_T} \}   \; .
\eeqa
Since only physical transverse states are included in the thermal
average, the longitudinal field part of the Hamiltonian has
only the vacuum to act on so there is no path integral for the
longitudinal field. The result is only transverse states need to be
periodic in time. There is nothing to say that the longitudinal
field has to be so, in contradiction with the usual formalism
where all bosonic fields are automatically taken to be periodic.

{}From the definition of the contour ordered propagator \cite{lands&weert},
periodic (anti-periodic) fields imply Kubo, Martin, Schwinger condition
(KMS) or periodic (anti-periodic) boundary condition for the propagators
therefore the transverse gauge propagator and scalar propagator are
periodic with period $-i\b$ but the propagator of the non-periodic
longitudinal field has no such periodicity. (Note that the trace
identity $\tr (AB) = \tr (BA)$, which we do not have here, is not
necessary for the transverse gauge propagator and the scalar
propagator to be periodic.)

Because of the absence of periodicity in $-i\b$, the longitudinal
propagator, of which the time-momentum representation
is given by \cite{james&land}
\beq D^L_{ab}(t,\vo k)_{ij} = i \d_{ab} {{k_i k_j} \over {\vo k}^2}
     \q (-k_3 \mbox{\rm Im}\,t) |\mbox{\rm Im}\, t|  \; ,
\label{longprop}
\eeq
when Fourier transformed using the full time range, i.e. between
$-i\b$ and $i\b$, into imaginary energy space has bosonic as well
as fermionic mode. It is now not obvious how energy is conserved in
this formalism. Remember in the usual case, at each interaction
vertex, take three particles interaction for example, one has
the following Euclidean time integral
\beq \int^\b_0 \rd \t \; \exp \{ i(p^0_1+p^0_2+p^0_3) \t \}
\eeq
where $p^0_i, i=1,2,3$ are the integral multiple $\p T$ valued energies
entering the vertex. For the case of three ``usual'' bosons, this
integral gives a kronecker-delta, because the sum of the energies is even,
which ensures energy conservation. For the case of a ``usual'' boson-fermions
interaction, because of charge conservation, there must be a pair of
fermions interacting with one ``usual'' boson, so the energy sum is
still even and again we have energy conservation. But in the case of this
new formalism, longitudinal gluon or photon can have either even or odd
integral multiple of $\p T$ as energy, so now the energy sum at
the vertex needs not be even therefore we may or may not get a
kronecker-delta to ensure energy conservation. The obvious question
to ask is how is energy conserved. We will show that this is
indeed a problem of this formalism when we study some one-loop
propagators in the abelian theory in the following sections.
Before one can do that, one will have to find out the
physical states to do thermal average over. This is due to
the fact that Gauss' law is lost when one chooses to work in TAG.
One finds the physical states by requiring
\beq \lan phys| \; \exp ( \intxs \; \o^a \l^a) |phys'\ran
     = \lan phys| phys'\ran \; ,
\eeq
where $\l^a$'s are the Gauss' laws and $\o^a$'s are just devices
introduced to sum over colours. Their actual values are unimportant
since what matters is the matrix elements of any power of Gauss' law
between physical states vanish. This will be dealt with in detail in
section \ref{searchO}.

\section{In search of $\co$}

\label{searchO}

In James and Landshoff's original paper, it was shown that because
of the Euclidean time range being finite (from 0 to $\b$), one cannot
form matrix elements between free incoming and/or outgoing states.
One is not allowed to switch off interactions at finite time without changing
the theory completely. So one has to find the physical states at a certain
time, say t=0, in order to find thermal averages. They advocate an
expansion of the form
\beq |phys \ran = \co |T \ran \; ,
\eeq
where $|T\ran$ are the free transverse gluon states and $\co$ is an
operator series
\beq \co = 1+ e o_1 + e^2 o_2 + e^3 o_3 + \cdots  \; ,
\eeq
so $|T\ran$ are the physical states when $e=0$.

The operators $o_i$ are obtained by requiring the matrix elements
of any power of Gauss's law between physical states to be zero
to make up for the loss of Gauss' law in TAG. We shall see that
it is a very weak condition to find physical states.
Gauss' law in a non-abelian pure gauge theory in TAG is
\beq \l^a (t,\vo x) = \dad^a (t,\vo x) - g f^{abc}
     {\vo A}^b(t,\vo x) \ncdot \dot {\vo A}^c(t,\vo x) \; .
\label{gaussnonabel}
\eeq
James and Landshoff found that it was not sufficient to apply Gauss's
law once but must do so repeatedly to find the $o_1$.
To simplify matters, we will study $\co$ in an abelian theory.
It turns out this can be solved completely in this case.

\subsection{$\co$ in Abelian Theories}

\label{searchOA}

In this section, we will study scalar electrodynamics but the same
also holds for QED as far as finding $\co$ is concerned.
In scalar electrodynamics, the simplest langrangian is given by
(including $\f^3$ or $\f^4$ type self-interaction will obviously not
affect $\co$)
\beq \cl = \intxs \Big \{(\del_\mu-ie A_\mu) \f \Big \}^\dagger
                      \Big \{(\del_\mu-ie A_\mu) \f \Big \}
           -{1\over 4} \intxs F^{\mu\nu}F_{\mu\nu}
\label{sedlangr}
\eeq
so Gauss' law in TAG is
\beq \dad -ie (\dot \f^\dagger \f-\f^\dagger \dot \f) =0 \; .
\eeq
{}From now on we will write
\beq \F = (\dot \f^\dagger \f-\f^\dagger \dot \f)  \; ,
\eeq
and before we solve for $\co$, we will split each $o_i$
into two parts
\beq o_i = o_i^A + o_i^H
\eeq
for the anti-Hermitian part and Hermitian part of $o_i$ respectively.

At order $e^1$, we have to solve
\beq \lsta [\dad, o_1^A] + \{\dad, o_1^H \} -i \F \rsta =0 \; .
\eeq
Here the states are free transverse photon states and free scalar
states. If we assume
\beq \lsta [\dad, o_1^A] -i \F \rsta =0 \; ,
\label{eqo1A}
\eeq
then we have, using the commutation relations in \cite{james&land,land},
the solution
\beq ^\l\!o_1^A = \intxs {1\over \Del^2} \da(x) \F (x) \; .
\label{o1A}
\eeq
The superscript $\l$ is to remind ourselves that this is the solution
of $o_1^A$ to using Gauss' law once. The remaining equation
\beq \lsta \{ \dad, o_1^H \} \rsta = 0 \; ,
\label{eqo1H}
\eeq
does not define $o_1^H$ very well. $o_1^H$ can be a term with no
$\da$ or product of more than one $\da$ or term with at least
one $\dad$. They all satisfy the above equation.
Applying Gauss' law repeatedly does not uniquely determine
$o_1^H$ although it does eliminate some possibilities. The same
can also be said about $o_1^A$. One can easily add a term to
\eref{o1A} and still satisfies \eref{eqo1A}. We will aim at finding
the simplest solution of $\co$ to the Gauss' law constraint so
we do not put an unknown term to \eref{o1A} in the hope that we might
be able to solve for it later. Also the simplest solution for
$o_1^H$ is zero. We will do this as long as we do not run into
equations that cannot be solved. In that case, it could be a sign that
extra terms are needed. It turns out that in an abelian theory such
as scalar electrodynamics or QED, it works. But in non-abelian theories,
it is much more difficult, and we will not go into this case in
this work. In our present case, we set
\beq ^\l\!o_1^H = 0 \; .
\label{o1H}
\eeq

Applying Gauss' law twice at order $e^1$, we have
\beqa \lsta \Big [\{\dad(x),\dad(y)\}, o_1^A \Big ] \fx&+&\fx
      \Big \{\{\dad(x),\dad(y)\}, o_1^H \Big \}  \nonum
      -i\{\F (x), \dad(y)\} \fx&-&\fx
      i\{\F (y), \dad(x)\} \rsta =0 \; .
\eeqa
One easily sees that using
\beq \Big [o_1^A, \{\dad(x),\dad(y)\} \Big ] =
     \Big \{[o_1^A,\dad(x)], \dad(y)\Big \} +
     \Big \{\dad(x), [o_1^A, \dad(y)]\Big \} \; ,
\eeq
the simplest solution at one $\l$ level \eref{o1A} and \eref{o1H}
also holds here. That is $^\l\!o_1^A =^{\l^2}\!o_1^A$ and
$^\l\!o_1^H = ^{\l^2}\!o_1^H$.

Perhaps one should note that
\beq [\F (x), \F(y) ] = 0 \; ,
\eeq
and is essentially the reason why an abelian theory is so much
simpler. The equivalence in QED is to replace $\F$ in \eref{o1A}
by $\bar \j \g^0 \j$
and we obviously have
\beq [\bar \j (x)\g^0 \j (x), \bar \j (y)\g^0 \j (y)] = 0 \; .
\eeq
It is now clear that the solution \eref{o1A} and
\eref{o1H} will satisfy the Gauss' law constraint at order $e^1$
no matter how many times it is applied. Henceforth the $\l$
superscript of $o_1$ will be dropped.

At order $e^2$, applying Gauss' law once gives
\beq \lsta [\dad,o_2^A] + \{\dad,o_2^H\} -i[\F,o_1^A]
     +o_1^\dagger \dad o_1 \rsta = 0 \; .
\eeq
The last two terms are vanishing and the remaining two terms
cannot cancel each other and have to vanish individually. So
\beq ^\l\!o_2=0
\eeq
is the simplest.

At the $\l^2$ level, after some rearrangement, we get
\beq \lsta \Big [\{\dad(x), \dad(y)\}, o_2^A\Big ]
    +\Big \{o_2^H-\half (o_1^A)^2, \{\dad(x), \dad(y)\} \Big \} \rsta =0\;.
\eeq
The simplest solutions at this level are then
\beq o_2^A =0\; , \mbox{\hskip 1.5cm}
     o_2^H = \half (o_1^A)^2 \; .
\eeq
These, in fact, satisfy all order $e^2$ Gauss' law constraints that is why
no superscript $\l^2$ has been given to them.
One can go on and it becomes clear that
\beq o_3^A = {1\over {3!}} (o_1^A)^3 \; , \mbox{\hskip 1.5cm}
     o_3^H = 0 \; .
\eeq
So we find, for scalar electrodynamics or QED,
\beqa \co \fx &=&\fx 1 + e o_1^A +{1\over {2!}} (e o_1^A)^2
                       + {1\over {3!}} (e o_1^A)^3 + \dots \nonum
          \fx &=&\fx \exp (e o_1^A)  \; .
\eeqa
In scalar electrodynamics $o_1^A$ is given by \eref{o1A}
whereas in QED, it is
\beq o_{1\; QED}^A = i \intxs {1\over \Del^2} \da(x)
     \bar \j(x) \g^0 \j(x) \; .
\eeq

So we find that $\co$ is unitary and therefore if the $\rsta$
is normalized so is $\co\rsta$. With these results, we can now
investigate the one-loop propagator.

\section{One-loop Propagator in the Abelian Theory}

\label{secprop}

In this section, we investigate the problem of energy
conservation in TAG in this formalism which has already been
allured to in the previous sections. We do that by looking at the
one-loop propagators in scalar electrodynamics. Obviously,
the propagator for the transverse gluon is energy conserving
at the one-loop level since both the scalar field and the
transverse propagator remain the same as in the usual formalism.
The same is also true for the one-loop scalar propagator
formed with a transverse gluon internal line.
So we start by studying the scalar propagator with a
longitudinal gluon internal line.

In this formalism, there are four diagrams that contribute
to the one-loop scalar propagator with a longitudinal gluon
line, $\cd^\f_l$. The diagram with no $o_1$ vertex, with one
$o^\dagger_1(-i\b)$ at the left vertex, with one $o_1(0)$ at the
right vertex and the one with two $o_1$'s i.e. with $o^\dagger_1(-i\b)$
at the left and $o_1(0)$ at the right vertex as shown in Fig. 1.
(The diagrams constructed from the $e^2$ term of
$\co$, that is those with two $o_1(0)$'s or with two
$o^{\dagger}_1(-i\b)$'s are zero because $D^L(t=0)=0$.)
We will let the Euclidean times $\t_1$ and $\t_2$ to be the time
integration variables at the left and at the right vertex
respectively.

Combining the four diagrams, we have the following expression
\beqa \cd^\f_l (s_1,s_2,\vo k)= e^2 \intqs \fx & & \fx \intps (2\pi)^3
      \d (\vo p+\vo q+\vo k) {1 \over {\vo q^2}}
      T^3\sum_{\klo,\kto,p_0} {1 \over {K_1^2 K_2^2 P^2}}
     \nonum
     \int^\b_0 \rd\t_1 \int^{\t_1}_0 \rd\t_2 \Big \{
         (\vo k^2-\vo p^2)^2
    \!\! &+&\!\! i(p_0+\klo) (\vo k^2-\vo p^2)\d(\b-\t_1)
                +i(p_0+\kto) (\vo k^2-\vo p^2) \d(\t_2)    \nonum
    \!\! &+&\!\! (\vo p^2+i p_0 i(\klo+\kto)+i\klo i\kto)
                 \d(\b-\t_1) \d(\t_2) \Big \} \nonum & &
     e^{ip_0(\t_1-\t_2)} e^{-i\klo(\t_1-\s_1)} e^{-i\kto(\s_2-\t_2)}
     D^L(t_1-t_2,\vo q)
\eeqa
where $\klo$ and $\kto$ are the incoming (from the left) and outgoing
energies (to the right) respectively, $K^2_i=(i k_{i0})^2-\vo k^2$,
$P^2=(ip_0)^2- \vo p^2$, and $t_i = -i\t_i \;, i=1,2$\ and $\s_i$'s
are the Euclidean external times (that is this propagator
starts at $s_1=-i\s_1$ and ends at $s_2=-i\s_2$). We are restricting
ourselves to the integration range in which $D^L$ will
give $\q(q_3)$. The other integration range will give the same
answer but with $\q(-q_3)$ so the sum is independent of the
direction of $q_3$.

We first convert the energy sum over $p_0$ into contour integral
and then use the relation
\beq \delta (\t) = \dbd \t \q (\t)
\label{del&therel}
\eeq
and partial integration over $\t$ to remove the delta functions.
The surface terms are vanishing either because
\beq D^L (0,\vo q) = 0 \; .
\label{zeroDLzero}
\eeq
or because
\beq \q (0) = 0 \; ,
\label{thetazero}
\eeq
which we choose to save on algebra and it has to hold
{\it for consistency} (see appendix \ref{Ad&q}) if one uses
\beq \int^\b_0 \rd\t \delta(\t) = 1 \; .
\eeq
One can of course choose the more conventional $\q(0)=\half$
but then one will have to use
\beq \int^\b_0 \rd\t \delta(\t) = \half \; .
\eeq
With this latter choice, the surface terms do not go away but
the final result is the same.

After some algebra, using
\beq T \sum_{p_0} \; \exp \{ip_0 (\t-\t') \} = \d(\t-\t')
\eeq
and
\beq \dbd {\t_1} \dbd {\t_2} D_L(t_1-t_2, \vo q) = -i \d(\t_1-\t_2)
     \; ,
\eeq
we get
\beqa \cd^\f_l (s_1,s_2,\vo k) = \fx && \fx e^2 \intqs \intps (2\pi)^3
      \d (\vo p+\vo q+\vo k) {1 \over {\vo q^2}}
      \sum_{s=\pm 1} {{s N(s \vmo p)} \over {2\vmo p}} i\q(q_3) \nonum
      \bigg \{ \fx &-&\fx T \sum_{\kto} {{(i\kto+s\vmo p)} \over {K_2^2}}
      \Big (
       {{e^{i\kto (\s_1-\s_2)}-e^{-i\kto \s_2+s\vmo p\s_1}} \over {i\kto-s\vmo
p}}
      \Big )
      \nonum \fx &-&\fx T \sum_{\klo} {{(i\klo+s\vmo p)} \over {K_1^2}}
      \Big (
       {{e^{i\klo (\s_1-\s_2)}-e^{i\klo \s_1+s\vmo p(\b-\s_2)}} \over
{i\klo-s\vmo p}}
      \Big )
        \nonum  \fx &+&\fx
       T^2\sum_{\klo,\kto} {{(s\vmo p+i\klo) (s\vmo p+i\kto)} \over {K_1^2
K_2^2}}
       \b \d_{\klo,\kto} e^{i\klo(\s_1-\s_2)} \bigg \}  \; ,
\eeqa
where $N(\vmo p)$ is the Bose-Einstein distribution function.
We see that in the first two terms, the last term within each round
brackets do not have time-translation invariance. This happens when the
poles at $\pm \vmo k$ are picked up by the contour integrals which
replace the energy sums. We now show that neither the one-loop
longitudinal propagator, $\cd^L$, has time-translation invariance.

We again have four diagrams to consider plus the tadpole graph.
(Again diagrams constructed with two $o_1(0)$'s or two
$o^\dagger_1(-i\b)$'s are zero. This time, this is due to the
product of the opposing theta functions $\q(k_3)\q(-k_3)$ coming
from the longitudinal propagators of the external lines.) Ignoring
the tadpole for the time being and with the same arrangement of the
$o_1$ vertices as in the previous case, Fig. 2, we get
\beqa \cd^L_{ij}(s_1,s_2,\vo k)=-e^2 \intqs \intps \fx && \fx (2\pi)^3
      \d(\vo p+\vo q+\vo k) {{k_i k_j} \over {(\vo k^2)}^2}
      T^2\sum_{p_0,q_0} {1 \over {P^2 Q^2}} \nonum
      \int^\b_{\s_1} d\t_1 \int^{\s_2}_0 d\t_2 \Big \{
     (\vo p^2-\vo q^2)^2  \fx &+&\fx
    i(p_0-q_0) (\vo p^2-\vo q^2) \big (\d(\b-\t_1) +\d(\t_2) \big )    \nonum
    +(\vo p^2+\vo q^2-2ip_0 iq_0) \d(\b-\t_1) \d(\t_2) \Big \} \fx &&\fx
    e^{-i(p_0+q_0)(\t_1-\t_2)} D^L(t_1-s_1,\vo k) D^L(s_2-t_2,\vo k)
    \; , \nonum & &
\label{longpropstart}
\eeqa
where $s_i=-i\s_i, i=1,2$ and again we take $\s_2> \s_1$.
We evaluate the energy sums by contour integrals, this requires us to
split the time integrals into various range and switching $p_0,q_0
\rightarrow -p_0,-q_0$ if necessary for convergence at infinity on the
complex $p_0$ and $q_o$ energy planes. After some cancellation between
the various graphs we get
\beqa \cd^L_{ij}(s_1,s_2,\vo k)= e^2 \intqs \intps \fx && \fx (2\pi)^3
      \d(\vo p+\vo q+\vo k) {{k_i k_j} \over {(\vo k^2)}^2} \sum_{r,s=\pm 1}
      {{r N(r \vmo p)} \over {2\vmo p}} {{s N(s \vmo q)} \over {2\vmo q}}
\q(k_3)
      \nonum (\vo p^2-\vo q^2)^2
      \bigg \{ {1\over {(r\vmo p+s\vmo q)^4}} \fx && \fx
              -{{2(\s_2-\s_1)} \over {(r\vmo p+s\vmo q)^3}}
              -{{(\s_2-\s_1)^3} \over {3(r\vmo p+s\vmo q)}}
              +{{e^{(r\vmo p+s\vmo q)(\s_2-\s_1)}} \over {(r\vmo p+s\vmo q)^4}}
              \nonum \fx & & \fx
              -{{e^{(r\vmo p+s\vmo q)\s_2}} \over {(r\vmo p+s\vmo q)^4}}
              -{{e^{(r\vmo p+s\vmo q)\s_1}} \over {(r\vmo p+s\vmo q)^4}}
      \bigg \} \; .
\label{longpropend}
\eeqa
The last two terms explicitly do not have time-translation invariance.

One can of course verify that this result is correct by evaluating the time
integrals first before doing the energy sums but more interestingly
a simpler check can be performed in the special case of $\s_1=\s_2=\s$.
In \eref{longpropstart}, we can now evaluate the energy sums and
there is no need to split up the time integrals into various range.
Furthermore, one can now use \eref{del&therel}, \eref{zeroDLzero} and
\eref{thetazero} to have some quick cancellations because the surface
terms can again be made to vanish. Doing it this way, one gets
\beq 2 (r\vmo p-s\vmo q)^2
     {{1-e^{(r\vmo p+s\vmo q)\s}} \over {(r\vmo p+s\vmo q)^2}}
\eeq
for the whole factor of the last two lines of \eref{longpropend} which agrees
with that in that equation when $\s_1=\s_2=\s$. We stress again that this
result is independent of the exact choice of the value of the theta
function at zero. We choose it to vanish at that point only for convenience.

One may wonder in this case, the terms without time-translation
invariance may be canceled by those coming from the tadpole graph.
Unfortunately, this is not the case. The contribution from the
tadpole, Fig. 3, preserves time-translation invariance and
is therefore energy conserving. It is
\beq \cd^L_{ij \; tadpole}(s_1,s_2,\vo k) = e^2 T\sum_{p_0} \intps
     {{k_i k_j} \over {\vo k}^2} {1 \over {P^2}}
     \int^\b_0 \rd \t D^L(t-s_1,\vo k) D^L(s_2-t,\vo k)  \; .
\eeq
The time integral give
\beq {{(\s_2-\s_1)^3} \over 6} \; .
\eeq
Therefore the complete one-loop longitudinal propagator has no
energy conservation.

\section{An Argument for Energy Conservation}

\label{puzzle}

In this section, we will give an argument for energy conservation
and therefore time-translation invariance of any thermal
N-point function in order to better explain the reason for the
counter examples of the one-loop propagators that we have
found above.

Thermal averages of an operator $Q$ expressed in the interaction picture
is
\beq \lan Q(t) \ran = Z^{-1} \sum_{phys} \lan phys| \; \exp(-\b H_{0I})
U(-i\b,t)
     Q_I(t) U(t,0) \; |phys \ran \; ,
\label{thermav1}
\eeq
where the sum is over physical states and we have chosen $t=0$ to be the
time when the interaction picture coincides with the Heisenberg picture.
Therefore a general N-point function in scalar electrodynamics has the form
\beq \lan A(t_1) \f(t_2) \dots A(t_n) \ran
     = Z^{-1} \sum_{phys} \lan phys| \; \exp(-\b H_{0I})
       B_n(t_1,t_2,\dots,t_n) \; |phys \ran \; ,
\eeq
where we have assumed $t_1 > t_2 > \dots > t_n$ and
\beq B_n(t_1,t_2,\dots,t_n)
     = U(-i\b,t_1) A_I(t_1) U(t_1,t_2)
     \f_I(t_2) U(t_2,t_3) \cdots A_I(t_n) U(t_n,0) \; .
\eeq
The fields here can be any combination of gauge fields and scalar fields
provided we have charge conservation.

In particular, the 2-point gauge function is
\beqa & & \lan A(t_1) A(t_2) \ran \nonum
      &=& Z^{-1} \sum_{phys} \lan phys| \; \exp(-\b H_{0I}) B_2(t_1,t_2)
          \; |phys\ran \\
      &=& Z^{-1} \sum_{phys} \lan phys| \; \exp(-iH \d)
         \exp(-\b H_{0I}) B_2(t_1+\d,t_2+\d) \exp(iH \d) \; |phys\ran \; .
\label{twopt}
\eeqa
The second line here is obtained by using the equation
\beq Q_I(t+\d) = \exp(iH_{0I}\d) \, Q_I(t)\, \exp (-iH_{0I}\d) \; ,
\eeq
for an operator $Q$ in the interaction picture.

Had we been working with a non-gauge theory, we could have inserted
two complete sets of states into \eref{twopt} and removed the
factors $\exp(\pm iH\d)$. Or equivalently, using the trace identity
$\tr (AB) = \tr (BA)$. But since in a gauge theory, part of the
Hilbert space is not physical and we are working in a ghostless
gauge, this trick is no longer available to us. Instead we argue
that since the physical states $|phys\ran$ can be expressed in terms of
energy eigenstates of the full Hamiltonian, the exponentials with
the full Hamiltonian $H$ at either end of the sum over physical states
of \eref{twopt} cancel each other. So we have
\beq \lan A(t_1) A(t_2) \ran =
     \lan A(t_1+\d) A(t_2+\d) \ran \; .
\eeq
The same also applies to the scalar two-point function.
Time translation invariance means energy conservation, therefore
the 2-point function is energy conserved in momentum space.
In other words, the energy coming in equals to the one
going out of the propagator. This obviously also works for
any N-point function so we should have total sum
of energies going into a graph to be zero.
This conclusion obviously contradicts what we have found
in section \ref{secprop}, so what is wrong?
To understand exactly what is going on, one will have to
study the longitudinal gauge field and the Hamiltonian
which are so constructed to yield the Vienna prescription
in the zero temperature longitudinal propagator. This leads
us to the next section.

\section{The Longitudinal Gauge Field and The Hamiltonian of
James and Landshoff's Formalism}

\label{explain}

We will point out the unusual features and peculiarities
of James and Landshoff's longitudinal field and Hamiltonian
in this section and give an explanation of the contradiction
that we have found previously.

In order to arrive at the Vienna prescription, the choice
of \cite{land}
\beq A_0= {\eta \over \del_3}{1 \over \Del^2} \da
\eeq
was made. The Hamiltonian formulation then forces the longitudinal
field to have four annihilation and creation operators
$q(\vo k), q^\dagger(\vo k), p(\vo k), p^\dagger(\vo k)$
as opposed to the usual two per field
\footnote{For the explicit expression of the longitudinal field written
in terms of these operators, we refer to \cite{james&land,land}.}.
They satisfy
\beq [q(\vo k), p^\dagger(\vo k')] = {\vo k}^2 (2\pi)^3
     \d (\vo k- \vo k') \; ,
\label{pqcom}
\eeq
and its Hermitian conjugate. All other commutation relations
between them are vanishing. Observe that in the usual case
of two operators per field, one has
\beq [a(\vo k), a^\dagger(\vo k')] = 2k_0 (2\pi)^3
     \d (\vo k- \vo k') \; .
\eeq
The change from $2k_0$ to $\vo k^2$ is unimportant here, it
is just a matter of dimensions. What is important is
the operators commute with their Hermitian conjugates which
is not the usual case. Therefore one can construct
states using the $q^\dagger$ and the $p^\dagger$ which
have zero norm. For example, the simplest zero norm states are
\beq q^\dagger (\vo k) |0\ran \mbox{\hskip 1cm and \hskip 1cm}
     p^\dagger (\vo k) |0\ran  \; .
\eeq
So the completeness of states is constructed by Dirac's
bra and ket state vectors not only of the usual form where
the bra is related to the ket by a Hermitian conjugate,
eg. $|\vo k\ran \lan \vo k|$, but also of the form where they are
not related by such a relation, eg.
$q^\dagger(\vo k')|\vo k\ran \lan \vo k|p(\vo k')$ \cite{land}.

The longitudinal part of the normal ordered free Hamiltonian in
terms of these operators is \cite{land}
\beq H^L_0 = \intks {{\q(k_3)} \over {\vo k}^2}
     \Big \{ p^\dagger(\vo k) p(\vo k) + {\eta \over k_3}
     \Big (p^\dagger(\vo k) q(\vo k)+q^\dagger(\vo k) p(\vo k)
     \Big ) \Big \}  \; ,
\eeq
so that we have by acting this on the simplest states above
\beqa H^L_0 q^\dagger(\vo k)|0\ran \fx &=& \fx
      \Big (p^\dagger(\vo k)
           +{\eta \over k_3} q^\dagger(\vo k) \Big ) |0\ran \\
      H^L_0 p^\dagger(\vo k)|0\ran \fx &=& \fx
      {\eta \over k_3} p^\dagger(\vo k) |0\ran  \; .
\eeqa
So the state created by $q^\dagger$ is not an energy
eigenstate of $H^L_0$. One {\it cannot} construct an energy eigenstate
by some combination of it with the state created by $p^\dagger$.
This is not a problem in the free theory because such
states are not physical either because of the Gauss' law
constraint or because of the zero norm. They live outside the
physical part of the Hilbert space. The problem comes
when one turns on the interaction. The physical Hilbert space
is now extended somewhat into the previously non-physical
part because the interaction Hamiltonian forms a
``bridge'' linking the two parts.

One can ask whether the above features have any effect
on the interacting theory in particular the energy eigenstates.
The answer is yes and we will now show it explicitly by
constructing energy eigenstates.

Since the part of the interacting Hamiltonian which links
the physical Hilbert space in the non-interacting theory
to itself produces nothing unusual, we will leave it out.
We concentrate only on that part which bridges the physical
and the non-physical part and only to first order in $e$.
That is we consider $H=H_0+e H_I$ where $H_I$ is
\beq H_I = -i\intxs {1 \over \Del^2} \da
           \Big ({\rd \over {\rd t}}-{\eta \over \del_3} \Big )
           \Big (\f^\dagger \dot \f-\dot \f^\dagger \f \Big ) \; .
\eeq
We treat $H_I$ as a small perturbation to the free
Hamiltonian $H_0$. The states at zero order in $e$ are the
free scalar and transverse photon states, i.e.
\beq H_0 |\f \; T\ran = E_{\f \; T}|\f \; T\ran \; .
\eeq
{}From order $e^1$ and above, the corrections to the free states are
constructed from the whole Hilbert space, i.e. the order $e^n$
correction to the perturbed free state $|\f \; T\ran$
is of the form
\beq |\f \; T\ran^{(n)}_{pert}
    = \sum_\a C^{(n)}_\a |\a\ran \; ,
\eeq
where the sum over $\a$ includes the integrals over momenta,
sum over spin and other quantum numbers etc. and goes over all
states in the Hilbert space. $H_I$ will automatically pick out
those states which become physical and leave out those that remain
non-physical. We choose to work out the energy eigenstates at t=0
when $A^L$ contains only $q$ and $q^\dagger$ and not $p$ and
$p^\dagger$.

To understand the following, it is only necessary to know
which matrix elements of $H_I$ is zero and which is not. Since $H_I$
is linear in $A_L$, at t=0, the only non-vanishing matrix elements
are those between a normal bra state and a suitable
\footnote{By suitable, we mean the bra and the ket must have the same
transverse photon state since $H_I$ does not have those and also the
scalar part which means not all scalar bra and ket combinations will
be non-vanishing, but this is not essential in the present discussion.
We would rather concentrate on the essential which is the $A_L$ part.}
ket state which contains one $p^\dagger (\vo k)$, and also of course
their complex conjugate.

{}From standard quantum mechanics perturbation theory, we know
that the first order perturbed energy is
\beq E^{(1)}_{\f \; T} = \lan \f \; T| H_I | \f\; T\ran
\eeq
(which is zero in this case) and the perturbed state to order $e^1$ is
\beq |\f \; T\ran_{pert} = |\f \; T\ran +e
     \sum_{\f'} \intks {{\q(k_3)} \over {\vo k^2}}
     {{q^\dagger(\vo k)|\f'\; T\ran
     \lan T\; \f'|p(\vo k) H_I |\f \; T\ran} \over
     {E_{\f\; T}-E_{\f'\; T}-\eta/k_3}}   \; .
\eeq
Notice that the energy being subtracted from that of the free state
in the denominator is that of $p^\dagger(\vo k)|\f'\; T\ran$ which
is an energy eigenstate of $H_0$. We have observed before that
the state with the $q^\dagger|0\ran$ is not an energy eigenstate
of $H_0$ so what we have constructed, is it really an energy
eigenstate of $H$ to order $e^1$? This we can verify readily.
Acting on the above energy eigenstate with $H$, we get to order $e^1$
\beqa H |\f \; T\ran_{pert} \fx &=& \fx E_{\f \; T}
     |\f \; T\ran_{pert}
    +e|\f \; T\ran\lan T \; \f|H_I|\f \; T\ran \nonum
    \fx & & \fx
    +e\sum_{\f'} \intks {{\q(k_3)} \over {\vo k^2}}
     {{p^\dagger(\vo k)|\f'\; T\ran
     \lan T\; \f'|p(\vo k) H_I |\f \; T\ran} \over
     {E_{\f\; T}-E_{\f'\; T}-\eta/k_3}}   \; .
\eeqa
Had we been working with a ``regular'' Hamiltonian and with ``regular''
states, the last term should not be there. Its presence is
due precisely to the fact that the states with $q^\dagger$ above
are not energy eigenstates of $H_0$. This means that we are not able
to construct energy eigenstates of the full Hamiltonian which
has all the peculiar features discussed above.

We can now return to the argument given in section \ref{puzzle}.
It seems that the physical states defined by Gauss' law cannot
be written as sums of energy eigenstates due to the fact that these
cannot even be constructed so that the factors $\exp(\pm iH \d)$ appeared
in the thermal averages of any N-point function due to the time
shift introduced deliberately cannot be removed. It follows that
\beq \lan A(t_1) A(t_2) \ran \neq
     \lan A(t_1+\d) A(t_2+\d) \ran \; .
\eeq
The inequality holds unfortunately also for the scalar field two-point
function and any N-point function. Therefore we do not have time
translation invariance in the propagators and in any N-point function.

\section{Conclusion}

In this paper, we have studied the TAG formalism initiated by James
and Landshoff in the abelian case. We find that in this case,
the $\co$ operator which links the transverse states to the
physical states can be solved completely and in fact, it is found
to exponentiate. The exponent is the order $e^1$ term of the
expansion of $\co$ in terms of the coupling. However, energy
conservation is missing in those one-loop propagators which
involve the longitudinal photon propagator and the $o_1$ vertex.
This lack of energy conservation or time translation invariance
can be traced back to the impossibility to construct energy
eigenstates due to the rather unconventional representation
of the longitudinal field and hence the Hamiltonian.
Those propagators that have energy conservation
at one-loop, i.e. the transverse photon propagator and the scalar
propagator with an internal transverse photon line, are so only
because they still do not know about the first order corrections
to the free states. Their higher-loop propagators will not have
energy conservation.

The source of the problem lies in the commutation relations
\eref{pqcom} between the four annihilation and creation
operators of the longitudinal field, which are, unfortunately,
necessary for the field to satisfy all the canonical commutation
relations and the free field equation of motion, $\ddot A_L=0$.
We do not understand why the above constraints of the Hamiltonian
formulation conspire in such a way so as to forbid itself from
having energy eigenstates.

What still holds in the formalism of James and Landshoff is that
the longitudinal field and propagator cannot be periodic.
In other words, KMS condition is not for TAG. One still
knows how to construct physical states at least in the abelian
theory since this depends only on the canonical commutation
relations. However, what remains to be understood is why we
cannot have a proper Hamiltonian formulation in TAG.

It must be said that the problem of energy conservation does
not arise in the real-time formalism where energy is conserved
everywhere. In that case, the time contour is chosen to start
from $-\infty$ on the complex time plane, goes to $t=\infty$,
and it ends eventually at $-\infty-i\b$ so any time shift
is easily absorbed by the infinity. Also, we only need the
physical states at $t=-\infty$ which are the energy eigenstates of
the free Hamiltonian, namely, the free transverse photon
and scalar states since we can switch off interactions then and
switch them on adiabatically.

\vskip 0.8cm
\noindent{\bf Acknowledgements}
\vskip 0.5cm

The author would like to thank P.V. Landshoff and J.C. Taylor for
discussions, N.H. Willis for some stimulating email exchanges,
and the referee for pointing out a mistake in an earlier version.
The author acknowledges financial support from the Leverhulme Trust.

\appendix

\section{Appendix}

\subsection{$\d$ and $\q$ Functions}

\label{Ad&q}

In this appendix, we will give some relations for manipulating
$\d$ and $\q$ functions. Given that $\q$ has the usual meaning
i.e.
\beqa \q(x) &=& 1  \mbox{\rm \hskip 2cm for $x > 0$,} \nonum
            &=& 0  \mbox{\rm \hskip 2cm for $x < 0$,}
\eeqa
then we have
\beq \int^\b_0 \rd \t \d(\t) = 1 \; .
\eeq
Then using $\dbd \t \q(\t) =\d(\t)$, we have
\beq \q(\b) - \q(0) = 1 \;,
\eeq
so $\q(0) = 0$ has to hold for consistency.

In general, one can choose
\beq \int^\b_0 \rd \t \d(\t) = \a
\eeq
where $\a$ is some number, then one must use $\q(0) = 1-\a$ in order
to be consistent.

\section {Figure Captions}

\begin{itemize}

\item[Fig. 1]Graphs contributing to part of the one-loop scalar
propagator with $\q(q_3)$. The loop consists of one
longitudinal and one scalar line.

\item[Fig. 2]Graphs contributing to part of the one-loop longitudinal
propagator with $\q(k_3)$. The loop consists of scalar lines.

\item[Fig. 3]The tadpole contribution to the one-loop longitudinal
propagator. This graph is energy conserved.

\end{itemize}


\begin{thebibliography}{99}

\bibitem{pei&wong}S. Peign\'e and S.M.H. Wong, Phys. Lett. B346 (1995)
332.

\bibitem{bra&pis} E. Braaten and R.D. Pisarski, Nucl. Phys. B337 (1990),
569.

\bibitem{kaj&kap} K. Kajantie and J. Kapusta, Ann. Phys. 160 (1985),
477.

\bibitem{hei&kaj&toi} U. Heinz, K. Kajantie and T. Toimela, Ann. Phys.
176 (1987), 218.

\bibitem{kap} J.I. Kapusta, Finite Temperature Field Theory
(Cambridge University Press, Cambridge, England, 1989).

\bibitem{reb1} A.K. Rebhan, Phys. Rev. D48 (1993), R3967.

\bibitem{reb2} A.K. Rebhan in Proceedings of the 3rd Workshop
on Thermal Field Theories and Their Applications, Banff,
Canada, August 1993, World Scientific, Edited by F.C. Khanna,
R.Kobes, G. Kunstatter and H. Umezawa, p.469.

\bibitem{reb3} A.K. Rebhan, Contributed talk at the NATO Advanced
Research Workshop ``Electroweak Physics and the Early Universe'',
22-25 March 1994, Sintra, Portugal, hep-ph/9404292.

\bibitem{baier&kal}R. Baier and O.K. Kalashnikov, Phys. Lett. B328
(1994) 450.

\bibitem{reb4}A.K. Rebhan, Nucl. Phys. B430 (1994) 319.

\bibitem{bra&nie}E. Braaten and Agustin Nieto, Northwestern University
preprint NUHEP-TH-94-18, August 1994.

\bibitem{nach}H. Nachbagauer, Z. Phys. C56 (1992) 407.

\bibitem{james&land}K. James and P.V. Landshoff, Phys. Lett. B251
(1989), 167.

\bibitem{gai&kreu}P. Gaigg and M. Kreuzer, Phys. Lett. B205 (1988) 530.

\bibitem{lei}G. Leibbrandt, Rev. Mod. Phys. 59 (1987) 1067.

\bibitem{land} P.V. Landshoff, Phys. Lett. B227 (1989), 427.

\bibitem{james1}K.A. James in Proceedings of the Workshop of Physical
and Nonstandard Gauges, Vienna, Austria, September 19-23, 1989, Edited
by P. Gaigg, W.E. Kummer, M. Schweda, Springer-Verlag, 1990, p.303.

\bibitem{james2}K.A. James, Z. Phys. C48 (1990) 169.

\bibitem{james3}K.A. James, Z. Phys. C49 (1991) 115.

\bibitem{lands&weert}N.P. Landsman and Ch.G.van Weert, Phys. Rep. 145
(1987) 141.

\end{thebibliography}
\end{document}